\documentclass[aps,onecolumn,amsmath,amssymb,amsfonts,showpacs]{revtex4}

\usepackage{graphics}
\usepackage{epsfig}
\usepackage{amsmath}
\usepackage{color}
\newcommand{\ket}[1]{\left | \, #1 \right\rangle}
\newcommand{\bra}[1]{\left \langle #1 \, \right |}

\begin{document}

\title{Steady state entanglement between hybrid light-matter qubits}
\date{\today}

 \author{Dimitris G. \surname{Angelakis}$^{1,2}$}
    \email{dimitris.angelakis@gmail.com}
\author{Stefano \surname{Mancini}$^{3}$}
\email{stefano.mancini@unicam.it}
\author{Sougato Bose $^{4}$}
\email{sougato.bose@googlemail.com}
\address{$^{1}$Centre for Quantum Computation, Department
of Applied Mathematics
 and Theoretical Physics, University of Cambridge,
 Wilberforce Road, CB3 0WA, UK}
\address{$^{2}$Science Department, Technical University of Crete, Chania,
Greece 73100}
\address{$^{3}$ Dipartimento di Fisica, Universit\`a di Camerino,
I-62032 Camerino, Italy}
  \address{$^{4}$Department of Physics and Astronomy, University
College London, Gower St., London WC1E 6BT, UK}
\begin{abstract}
We study the case of two polaritonic qubits localized in two
separate cavities coupled by a fiber/additional cavity. We show that
surprisingly enough, even a coherent classical pump in the
intermediate cavity/fiber can lead to the creation of entanglement
between the two ends in the steady state. The stationary nature of
this entanglement and its survival under dissipation opens
possibilities for its production under realistic laboratory
conditions. To facilitate the verification of the entanglement in an
experiment we also construct the relevant entanglement witness
measurable by accessing only a few local variables of each
polaritonic qubit.
\end{abstract}


\maketitle

\section{Introduction}

Recently, there has been a growing interest in exploiting
a certain class of  coupled hybrid light-matter systems, namely
coupled cavity polaritonic systems, for various purposes such as for
realizing schemes for quantum computation
\cite{angelakis-ekert04,angelakis-kay07a}, for communication
\cite{angelakis-bose07b} and for simulations of quantum many-body
systems \cite{angelakis-bose06b,hartmann,greentree,cpsun,fazio,yamamoto,
myungshik-agarwal,myungshik}.  These cavity-atom polaritonic
excitations are different from propagating polaritonic excitations
in atomic gases and exciton-photon polaritons in solid state
systems \cite{lukin-yamamoto}. This area is also distinct from those
using hybrid light-matter systems in quantum computing where only
the matter system (such as an atom or an electron) acts as the
qubit. In the latter case the qubits are atoms and light is used
exclusively as a connection bus between them
\cite{barrett,beige,munro,serafini,MW05,MB04,parkins}. Promising
schemes to produce steady state entanglement between atoms in
distinct cavities have also been proposed \cite{parkins}. In these
ground states of atoms have been used in order to circumvent
decoherence due to spontaneous emission. In addition to auxiliary
atomic levels, external driving fields as well as an unidirectional
coupling between cavities are required. In polaritonic coupled
cavity systems on the other hand, the localized mixed light-matter
excitations, or polaritons, allow for the identification of qubits
that possess the easy manipulability and measurability of atomic
qubits, while also being able to naturally interact whereas separated
by distances over which photons can be exchanged between them.
Motivated by the rapid experimental progress in Cavity Quantum
Electrodynamics and the ability to couple distinct cavities in a
variety of systems \cite{cqed,blockade,toroid,noda,trupke}, the
realization of a system that could produce verifiable, steady state
entanglement between two polaritonic qubits in currently realistic
laboratory conditions would be extremely interesting. In that case
the decoherence emerging from the photonic losses due to the mixed
nature of the polaritons, in addition to that from atomic
spontaneous emission, will need to be controlled. Therefore,
apriori one may not expect a completely stationary entanglement of
two polaritons unless the unavoidable loss of coherence due to both
channels can somehow be ``re-injected'' into the system.

Here we show that even under strong dissipation in both the atomic
and photonic parts, it is still possible to deterministically
entangle two such polaritonic qubits. More precisely, we study the
case of two polaritonic qubits coupled by a fiber/additional cavity
and show that surprisingly enough, even a coherent {\em classical
pump} can lead to the creation of entanglement between them in the
steady state. The stationary nature of this entanglement should make
easier its experimental  verification. To this end we also provide
a relevant operator (an ``entanglement witness" \cite{witness})
measurable by only measuring local variables of each polariton.


\section{The Model}

The Hamiltonian describing an array of $N$ identical atom-cavity systems is the sum of
 the free light and dopant parts and the internal photon and dopant
 couplings
\begin{eqnarray}
H^{free}&=&\omega_{d}\sum_{k=1}^N a_k^\dagger a_k+\omega_{0}\sum_{k=1}^N
|e\rangle_{k} \langle e|, \\
H^{int}&=&g \sum_{k=1}^N(a_k^\dagger\,|g\rangle_{k}\langle e|
+a_k |e\rangle_{k}\langle g|).
\end{eqnarray}
Here $a_k,a_k^{\dag}$ are the photonic field operators localized in
the $k$-th system and $|e\rangle_k,|g\rangle_k$ are the excited and
ground state of the dopant in the $k$-th system. Moreover, $g$ is
the light-atom coupling strength and $\omega_{d}$($\omega_0$) the
photonic(atomic) frequencies respectively ($\hbar=1$ throughout the
paper). The $H^{free}+H^{int}$ Hamiltonian can be diagonalized in a
basis of mixed photonic and atomic excitations, called {\it
polaritons}. On resonance between atom and cavity, the polaritons
are created by operators
$P_{k}^{(\pm,n)\dagger}=\ket{n\pm}_k\bra{g,0}$. The states
$\ket{n\pm}_k=(\ket{g,n}_k\pm \ket{e,n-1}_k)/\sqrt2$ are the
polaritonic states (also known as dressed states) with energies
$E^{\pm}_{n}=n\omega_{d}\pm g\sqrt{n}$ and $\ket{n}_k$ denotes  the
$n$-photon Fock state of the $k$-th cavity.

It has been shown that in an array of these atom-cavity systems
the addition of a hopping photon term
$\propto \sum_j(a^{\dagger}_{j}a_{j+1}+a_{j}a^{\dagger}_{j+1})$,
leads to a polaritonic Mott phase where a maximum of
one excitation per site is allowed \cite{angelakis-bose06b}. This
originates from the repulsion due to the photon blockade effect
\cite{blockade}. In this Mott phase, the system's Hamiltonian in the interaction picture results
\begin{equation}
H_{I}=J \sum_{k}\left(P_{k}^{(-,1)\dagger}P_{k+1}^{(-,1)}+
P_{k}^{(-,1)}P_{k+1}^{(-,1)\dagger}\right),
\end{equation}
where $J$ is the coupling due to photon hopping from cavity to
cavity. Since double or more occupancy of the sites is prohibited,
one can identify $P_{k}^{(-,1)\dagger}$ with
$\sigma^{\dag}_k=\sigma^x_k+i\sigma^y_k$, where $\sigma^x_k$,
$\sigma^y_k$ and $\sigma^z_k$ stand for the usual Pauli operators.
The system's Hamiltonian then becomes the standard $XY$ model of
interacting spin qubits with spin up/down corresponding to the
presence/absence of a polariton \cite{angelakis-bose06b}.

Let us now consider a linear chain of three
coupled cavities with the two extremal ones doped with a two level
system as shown in Fig.1(a). Alternatively, as the central cavity in
any case is undoped, one can simply replace it with an optical fiber
of short length (so that the distance is greatly increased but the
fiber still supports a single mode of frequency near those of the
two cavities), which simplifies the setting even further, as shown
in Fig.1(b). For the purposes of description, we will use the three
cavity setting remembering that everything applies to the case of
two cavities linked by a fiber. The fact that a classical field can
drive (i.e., pump energy into) the central cavity in a three cavity
setting (as also shown in Fig.1(a)) is replaced in the fiber setting
by a coupler feeding light into the cavity (as also shown in
Fig.1(b)).

\begin{figure}
\vspace{-2.0cm}
\begin{center}
\includegraphics[width=0.6\textwidth]{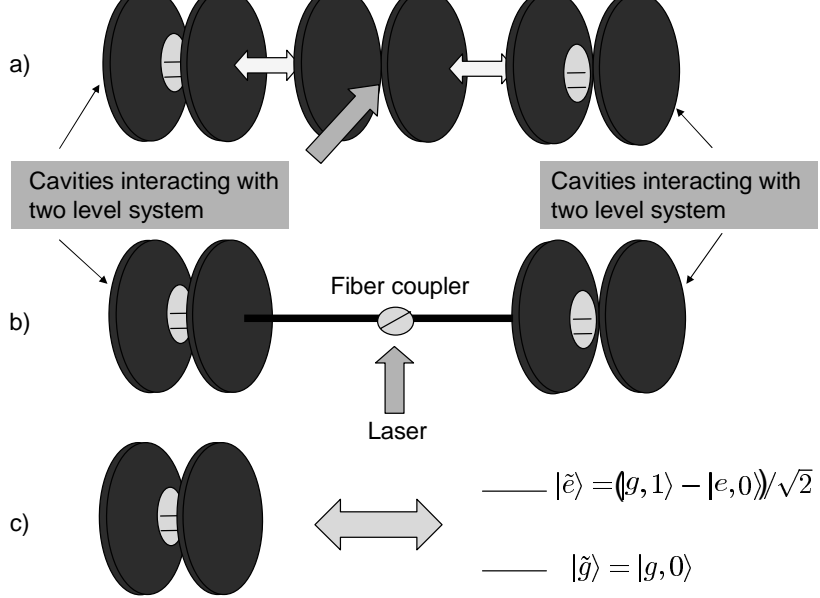}
\end{center}
\vspace{-2.0cm}
 \caption{\label{scheme} The system under consideration. a) The
cavities are coupled through direct photon hopping. b) The cavities
are coupled through a fiber. The extremal cavities in each
configuration are interacting with a two level system that could be
an atom or a quantum dot depending the implementation technology
used. c) The photon blockade allows for the ground and first dressed
states of each atom-cavity system  to be treated as a (polaritonic)
qubit.}
\end{figure}

Let  $\sigma_{j}^{\dagger}=|1-\rangle_{j}
\langle g, 0|$ be the polaritonic spin operators for the end cavities
(the index $j=1,2$ labels the two end cavities) and $a$, $a^{\dag}$
the field operators of the central empty cavity. Since the latter is not doped,
there the field operators play the role of polariton operators and they couple to polariton
operators of the ends cavities. Moreover, assuming that the central cavity (or fiber) is driven,
the Hamiltonian describing the system dynamics will be
\begin{eqnarray}
H=J\sum_{j=1}^{2}\left(\sigma_ja^{\dag}+\sigma_j^{\dag}a\right)-\Delta
a^{\dag}a +\alpha a^{\dag}+\alpha^*a,
\end{eqnarray}
where $\Delta=\omega_{mid}-\omega_{pol}$ is the detuning between the
central cavity mode of frequency $\omega_{mid}$ and the polaritons
frequency $\omega_{pol}=\omega_0-g$. Furthermore, $\alpha$ is the
product of the coupling of the driving field to the central cavity
field (say $G$) and the amplitude of the driving radiation field
(say $\tilde\alpha$). We also assume that $\Delta$ is much smaller
than the atom-light coupling in each of the outer cavities, so that
only the ground level $|\tilde{g}\rangle=|g,0\rangle$ and first
excited level $|\tilde{e}\rangle=(|g,1\rangle-|e,0\rangle)/\sqrt{2}$
of the polaritons are involved (i.e., the polaritons are still good
as qubits).

Suppose that the polaritons decay with the same rate $\gamma$ (this
is the effective decay rate of the polariton due to both the decay
of the cavity field and the atomic excited state), and the cavity
radiation mode with rate $\kappa$. The quantum Langevin equations
describing the dynamics will be \cite{qnoise}
\begin{eqnarray}
\dot{\sigma}_j&=&iJ
a\sigma_j^{z}-\gamma\sigma_j+\sqrt{2\gamma}\sigma_j^{in},\qquad\quad
j=1,2\label{eqssigma}\\
\dot{a}&=&i\Delta a-iJ\left(\sigma_1+\sigma_2\right)-i\alpha-\kappa
a+\sqrt{2\kappa}a^{in},\label{eqa}
\end{eqnarray}
where the superscript \emph{in} denotes the vacuum noise operators.

If $\kappa\gg J$ the radiation mode can be adiabatically eliminated
in such a way that
\begin{equation}
a\approx\frac{J}{\Delta+i\kappa}\left(\sigma_1+\sigma_2\right)+\frac{\alpha}{\Delta+i\kappa}
+i\frac{\sqrt{2\kappa}}{\Delta+i\kappa}a^{in}. \label{light1}
\end{equation}
Moreover, if the quantities $J/(2\sqrt{\kappa})$ and $\alpha/(2\sqrt{\kappa})$ are large
compared to the amplitude standard deviation of the fluctuating vacuum field, the last term  in
Eq.(\ref{light1}) can be neglected and
\begin{equation}
a\approx\frac{J}{\Delta+i\kappa}\left(\sigma_1+\sigma_2\right)+\frac{\alpha}{\Delta+i\kappa}.
\label{aaprox}
\end{equation}

Inserting Eq.(\ref{aaprox}) into Eqs.(\ref{eqssigma}), we
get\footnote{The term $a\sigma_j$ in Eqs.(\ref{eqssigma}) is
considered as to be  $(a\sigma_j+\sigma_j a)/2$.}
\begin{eqnarray}
\dot{\sigma}_1&=&i\frac{J^2}{\Delta+i\kappa} \sigma_2\sigma_1^{z}
+i\frac{J\alpha}{\Delta+i\kappa}\sigma_1^{z}
-\gamma\sigma_1+\sqrt{2\gamma}\sigma_1^{in},
\label{eqssigma1}\\
\dot{\sigma}_2&=&i\frac{J^2}{\Delta+i\kappa} \sigma_1\sigma_2^{z}
+i\frac{J\alpha}{\Delta+i\kappa}\sigma_2^{z}
-\gamma\sigma_2+\sqrt{2\gamma}\sigma_2^{in}, \label{eqssigma2}
\end{eqnarray}
corresponding to an effective Hamiltonian for polaritons of the type
\begin{eqnarray}
H_{eff}=\Re\left[\frac{J^2}{\Delta+i\kappa}\right]
\left(\sigma_1\sigma_2^{\dag}+\sigma_1^{\dag}\sigma_2\right)
+\frac{J\alpha}{\Delta+i\kappa}\left(\sigma_1^{\dag}+
\sigma_2^{\dag}\right)
+\frac{J\alpha^*}{\Delta-i\kappa}\left(\sigma_1+\sigma_2\right).
\end{eqnarray}
We are using $\Re$ and $\Im$ to denote the real and imaginary part respectively.

The dynamics of the polaritons can now be described by the master
equation \cite{qnoise}
\begin{eqnarray}
\dot{\rho} = -i\left[H_{eff},\rho\right] +\sum_{j=1}^{2} L_j\rho
L_j^{\dag}-\frac{1}{2}\left\{L_j^{\dag}L_j,\rho\right\}, \label{me}
\end{eqnarray}
where $L_j=\sqrt{2\gamma}\sigma_j$ are the Lindblad operators.


\section{Steady State Entanglement}

At the steady state Eq.\eqref{me} becomes
\begin{eqnarray}
0&=&-i\zeta\left[\sigma_1\sigma_2^{\dag}+\sigma_1^{\dag}\sigma_2,\rho\right]
-i\xi\left[\sigma_1^{\dag}+\sigma_2^{\dag},\rho\right]
-i\xi^*\left[\sigma_1+\sigma_2,\rho\right]\nonumber\\
&&+2\sigma_1\rho\sigma_1^{\dag}-\sigma_1^{\dag}\sigma_1\rho-\rho\sigma_1^{\dag}\sigma_1
+2\sigma_2\rho\sigma_2^{\dag}-\sigma_2^{\dag}\sigma_2\rho-\rho\sigma_2^{\dag}\sigma_2,
\label{meexp}
\end{eqnarray}
where $\zeta=\Re[J^2/\gamma(\Delta+i\kappa)]$ and $\xi=\alpha
J/\gamma(\Delta+i\kappa)$.

The steady state solution of Eq.(\ref{meexp}) can be found by
writing the density operator and the other operators in a matrix
form, in the basis
$\mathbf{B}=\{|\tilde{e}\rangle_{1}|\tilde{{e}}\rangle_{2},|\tilde{g}\rangle_{1}|\tilde{{e}}\rangle_{2},
|\tilde{e}\rangle_{1}|\tilde{{g}}\rangle_{2},|\tilde{g}\rangle_{1}|\tilde{{g}}\rangle_{2}
\}$. Let us parametrize the density operator as
\begin{equation}
\rho= \left(\begin{array}{cccc} {\cal A}&{\cal B}_{1}+i{\cal B}_{2}&
{\cal C}_{1}+i{\cal C}_{2}&{\cal D}_{1}+i{\cal D}_{2}
\\
{\cal B}_{1}-i{\cal B}_{2}&{\cal E}& {\cal F}_{1}+i{\cal
F}_{2}&{\cal G}_{1}+i{\cal G}_{2}
\\
{\cal C}_{1}-i{\cal C}_{2}&{\cal F}_{1}-i{\cal F}_{2}& {\cal
H}&{\cal I}_{1}+i{\cal I}_{2}
\\
{\cal D}_{1}-i{\cal D}_{2}&{\cal G}_{1}-i{\cal D}_{2}& {\cal
I}_{1}-i{\cal I}_{2}&1-{\cal A}-{\cal E}-{\cal H}
\end{array}\right),
\label{rhoexp}
\end{equation}
where the matrix elements also respect the requirement that ${\rm
Tr}\{\rho\}=1$.
The matrix representation of the other operators
comes from
\begin{equation}
\sigma_{1}= \left(\begin{array}{cccc} 0&0&0&0
\\
1&0&0&0
\\
0&0&0&0
\\
0&0&1&0
\end{array}\right)\,,
\quad \sigma_{2}= \left(\begin{array}{cccc} 0&0&0&0
\\
0&0&0&0
\\
1&0&0&0
\\
0&1&0&0
\end{array}\right)\,.
\label{sigmaexp}
\end{equation}
By using matrices (\ref{rhoexp}) and (\ref{sigmaexp}) in the r.h.s.
of Eq.(\ref{meexp}), we get a single
complex matrix $M$ which must be equal to zero. Then, equating to
zero the entries of $M$ we get a set of equation for the entries of
$\rho$. Since $M$ is Hermitian we can consider
\begin{eqnarray}
M_{jj}&=&0,\quad j,k=1,2,3,4\\
{\Re}\{M_{jk}\}&=&0,\quad k>j\\
{\Im}\{M_{jk}\}&=&0,\quad k>j
\end{eqnarray}
so to have a set of $16$ linear equations. They are not all
independent because of the 15 unknown parameters $({\cal A}, {\cal
B}_1, {\cal B}_2,{\cal C}_1,{\cal C}_2,{\cal D}_1,{\cal D}_2,{\cal
E},{\cal F}_1,{\cal F}_2,{\cal G}_1,{\cal G}_2,{\cal H},{\cal
I}_1,{\cal I}_2)$.
Explicitly the set of equations results
\begin{eqnarray}\label{rhoss}
-4{\cal A}+2\xi_2{\cal B}_1-2\xi_1{\cal B}_2+2\xi_2{\cal
C}_1-2\xi_1{\cal
C}_2&=&0, \nonumber\\
-\xi_2{\cal A}-3{\cal B}_{1}-\zeta{\cal C}_2+\xi_2{\cal
D}_1-\xi_1{\cal D}_2+\xi_2{\cal E}+\xi_2{\cal F}_1
-\xi_1{\cal F}_2&=&0, \nonumber\\
\xi_1{\cal A}-3{\cal B}_{2}+\zeta{\cal C}_1+\xi_1{\cal
D}_1+\xi_2{\cal D}_2-\xi_1{\cal E}-\xi_1{\cal F}_1
-\xi_2{\cal F}_2&=&0, \nonumber\\
-\xi_2{\cal A}-\zeta{\cal B}_2-3{\cal C}_{1}+\xi_2{\cal
D}_1-\xi_1{\cal D}_2+\xi_2{\cal F}_1
+\xi_1{\cal F}_2+\xi_2{\cal H}&=&0, \nonumber\\
\xi_1{\cal A}+\zeta{\cal B}_1-3{\cal C}_{2}+\xi_1{\cal
D}_1+\xi_2{\cal D}_2-\xi_1{\cal F}_1
+\xi_2{\cal F}_2-\xi_1{\cal H}&=&0, \nonumber\\
-\xi_1{\cal B}_{2}-\xi_1{\cal B}_{2}-\xi_2{\cal C}_{1}-\xi_1{\cal
C}_{2}-2{\cal D}_{1} +\xi_2{\cal G}_{1}+\xi_1{\cal G}_{2}+\xi_2{\cal
I}_{1}+\xi_1{\cal
I}_{2}&=&0, \nonumber\\
\xi_1{\cal B}_{1}-\xi_2{\cal B}_{2}+\xi_1{\cal C}_{1}-\xi_2{\cal
C}_{2}-2{\cal D}_{2} -\xi_1{\cal G}_{1}+\xi_2{\cal G}_{2}-\xi_1{\cal
I}_{1}+\xi_2{\cal
I}_{2}&=&0, \nonumber\\
2{\cal A}-2\xi_2{\cal B}_1+2\xi_1{\cal B}_2-2{\cal E}-2\zeta{\cal
F}_2+2\xi_2{\cal G}_1-2\xi_1{\cal G}_2
&=&0, \nonumber\\
-\xi_2{\cal B}_{1}+\xi_1{\cal B}_{2}-\xi_2{\cal C}_{1}+\xi_1{\cal
C}_{2} -2 {\cal F}_{1}+\xi_2{\cal G}_{1}-\xi_1{\cal G}_{2}
+\xi_2{\cal I}_{1}-\xi_1{\cal I}_{2}&=&0,\nonumber\\
\xi_1{\cal B}_{1}+\xi_2{\cal B}_{2}-\xi_1{\cal C}_{1}-\xi_2{\cal
C}_{2} +\zeta{\cal E} -2 {\cal F}_{2}+\xi_1{\cal G}_{1}+\xi_2{\cal
G}_{2} -\zeta{\cal H}
-\xi_1{\cal I}_{1}-\xi_2{\cal I}_{2}&=&0,\nonumber\\
-\xi_2{\cal A}+2 {\cal C}_{1}-\xi_2{\cal D}_{1}+\xi_1{\cal
D}_{2}-2\xi_2{\cal E} -\xi_2{\cal F}_{1}-\xi_1{\cal F}_{2}
-{\cal G}_{1}-\xi_2{\cal H}+\zeta{\cal I}_{2}&=&-\xi_2,\nonumber\\
\xi_1{\cal A}+2 {\cal C}_{2}-\xi_1{\cal D}_{1}-\xi_2{\cal
D}_{2}+2\xi_1{\cal E} +\xi_1{\cal F}_{1}-\xi_2{\cal F}_{2}
-{\cal G}_{2}+\xi_1{\cal H}-\zeta{\cal I}_{1}&=&\xi_1,\nonumber\\
2{\cal A}-2\xi_2{\cal C}_1+2\xi_1{\cal C}_2 +2\zeta{\cal F}_2-2{\cal
H}+2\xi_2{\cal I}_1-2\xi_1{\cal I}_2&=&0,
\nonumber\\
-\xi_2{\cal A}+2 {\cal B}_{1}-\xi_2{\cal D}_1+\xi_1{\cal
D}_2-\xi_2{\cal E} -\xi_2{\cal F}_1+\xi_1{\cal F}_2+\zeta{\cal
G}_2-2\xi_2{\cal H}-{\cal
I}_1&=&-\xi_2,\nonumber\\
\xi_1{\cal A}+2 {\cal B}_{2}-\xi_1{\cal D}_1-\xi_2{\cal
D}_2+\xi_1{\cal E} +\xi_1{\cal F}_1+\xi_2{\cal F}_2-\zeta{\cal
G}_1+2\xi_1{\cal H}-{\cal
I}_2&=&\xi_1,\nonumber\\
2{\cal E}-2\xi_2{\cal G}_1+2\xi_1{\cal G}_2+2{\cal H}-2\xi_2{\cal
I}_1+2\xi_1{\cal I}_2&=&0, \label{seteqs}
\end{eqnarray}
where $\xi_1=\Re\{\xi\}$ and $\xi_2=\Im\{\xi\}$.

Solving analytically the above set of equations we obtain for $\xi_2=0$
\begin{eqnarray}
\mathcal{A}&=&\frac{\xi_1^4}{d}\,,\quad \mathcal{B}_1=0\,,\quad
\mathcal{B}_2=-\frac{\xi_1^3}{d}\,,\quad \mathcal{C}_1=0\,,\quad
\mathcal{C}_2=-\frac{\xi_1^3}{d}\,,\quad
\mathcal{D}_1=-\frac{\xi_1^2}{d}\,,\quad
\mathcal{D}_2=\zeta\frac{\xi_1^2}{d}\,,\quad
\mathcal{E}=\frac{\xi_1^2+\xi_1^4}{d}\,,\nonumber\\
\mathcal{F}_1&=&\frac{\xi_1^2}{d}\,,\quad \mathcal{F}_2=0\,,\quad
\mathcal{G}_1=-\zeta\frac{\xi_1}{d}\,,\quad
\mathcal{G}_2=-\frac{\xi_1+\xi_1^3}{d}\,,\quad
\mathcal{H}=\frac{\xi_1^2+\xi_1^4}{d}\,,\quad
\mathcal{I}_1=-\zeta\frac{\xi_1}{d}\,,\quad
\mathcal{I}_2=-\frac{\xi_1+\xi_1^3}{d}\,,
\end{eqnarray}
where
\begin{equation}
d=\zeta^2+(1+2\xi_1^2)^2.
\end{equation}
Notice that  for $\xi_1=0$ we have formally analogous solutions
that lead to the same physical result, hence they are not reported.

Now that we know the stationary density matrix, we can use the
concurrence as measure of the degree of entanglement \cite{Woot}
\begin{equation}
C(\rho)=\max\left\{0,\lambda_1-\lambda_2-\lambda_3-\lambda_4\right\},
\end{equation}
where $\lambda_i$'s are, in decreasing order, the nonnegative square
roots of the moduli of the eigenvalues of $\rho\tilde\rho$ with
\begin{equation}
\tilde\rho=\left(\sigma_{1}^{y}\sigma_{2}^{y}\right)\rho^*\left(\sigma_{1}^{y}\sigma_{2}^{y}\right),
\end{equation}
and $\rho^*$ denotes the complex conjugate of $\rho$.
With respect to the basis $\mathbf{B}$ it results
\begin{equation}
\tilde\rho= \left(\begin{array}{cccc}
1-{\cal A}-{\cal E}-{\cal H}&-{\cal I}_{1}-i{\cal I}_{2}&
-{\cal G}_{1}-i{\cal G}_{2}&{\cal D}_{1}+i{\cal D}_{2}\\
-{\cal I}_{1}+i{\cal I}_{2}&{\cal H}& {\cal F}_{1}+i{\cal F}_{2}&-{\cal C}_{1}-i{\cal C}_{2}\\
-{\cal G}_{1}+i{\cal G}_{2}&{\cal F}_{1}-i{\cal F}_{2}&{\cal E}&-{\cal B}_{1}-i{\cal B}_{2}\\
{\cal D}_{1}-i{\cal D}_{2}&-{\cal C}_{1}+i{\cal C}_{2}&
-{\cal B}_{1}+i{\cal B}_{2}& {\cal A}
\end{array}\right),
\label{rhotil}
\end{equation}

\begin{figure}
\vspace{-2.0cm}
\begin{center}
\includegraphics[width=0.6\textwidth]{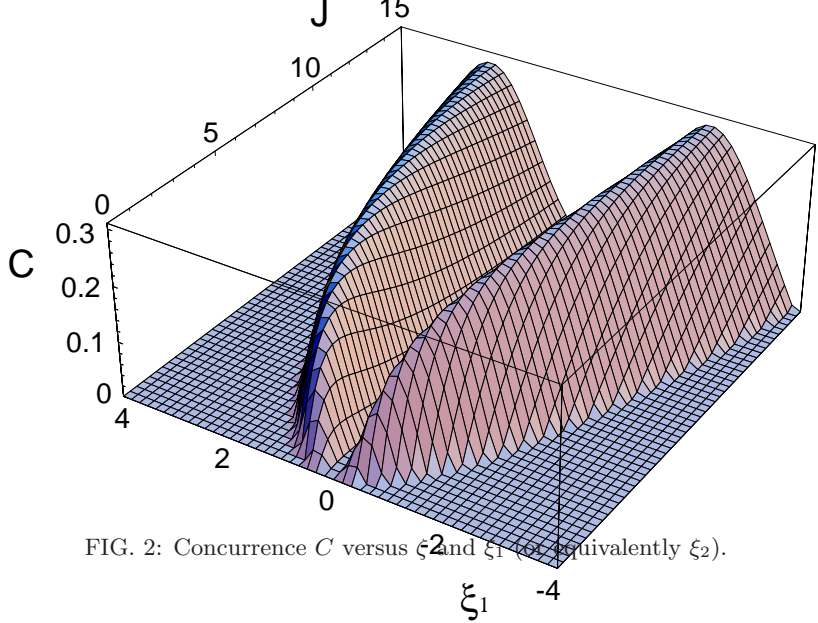}
\end{center}
\vspace{-2.0cm}
\caption{\label{concurrence}  Concurrence $C$ versus
$\zeta$ and $\xi_1$ (or equivalently $\xi_2$). }
\end{figure}

In Fig.\ref{concurrence} we show the concurrence as a function of
$\zeta$ and $\xi_{1}$ (the cases $\xi_1=0$ and $\xi_2=0$
give the same numerical results for the concurrence).
Notice that by increasing $\zeta$, the
concurrence increases quite slowly, and a maximum amount of
entanglement is approximately $0.3$ for $\zeta=10$ and
$\xi_{1}=2.135$. This is similar to the amount of stationary
entanglement achievable with an effective interaction of the kind
$\sigma_1^z\sigma_2^z$ when combined with an intricate feedback and
cascading \cite{MW05}.

\begin{figure}
\vspace{-2.0cm}
\begin{center}
\includegraphics[width=0.6\textwidth]{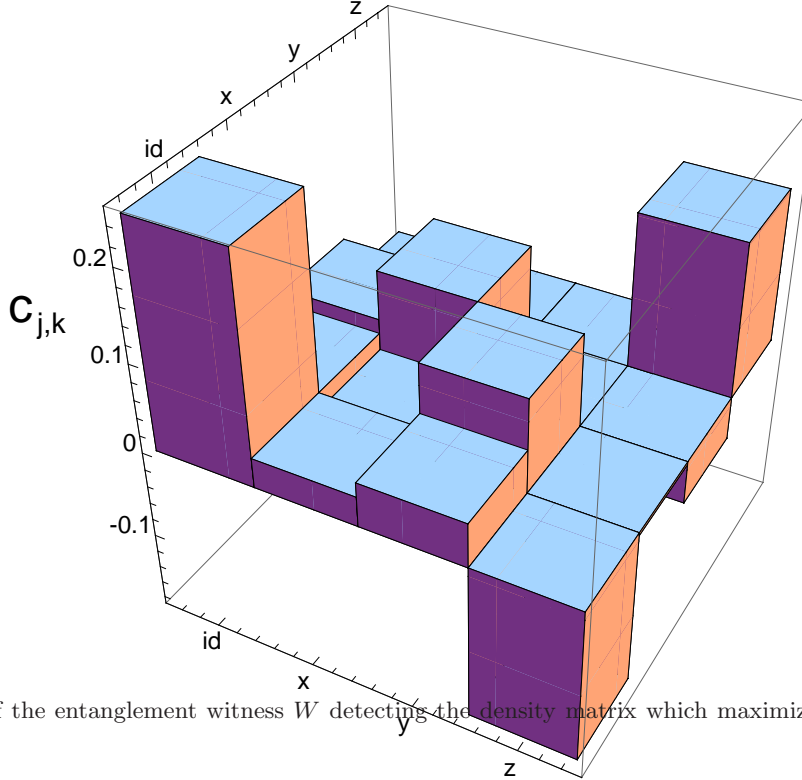}
\end{center}
\vspace{-2.0cm} \caption{\label{witness}
Elements $c_{j,k}$ of the entanglement witness $W$
detecting the density matrix which maximizes the concurrence in this
system.}
\end{figure}

One could try to employ entanglement
witnesses to detect this entanglement \cite{witness}.
A witness can be constructed from the density matrix
corresponding to the maximum value of the concurrence.
This would be a traceclass operator $W$ in the Hilbert space of the two polaritonic qubits
 such that ${\rm Tr}[W\rho]\ge 0$ for all separable states while ${\rm Tr}[W\rho]< 0$ for the considered entangled state. The form of such a
witness in the Pauli decomposition results
\begin{equation}
W=\sum_{j,k=id,x,y,z}
c_{j,k}\,\sigma_1^{j}\otimes \sigma_2^{k},
\end{equation}
 with $\sigma^{id}=I$.
In Fig.\ref{witness} we show the coefficients $c_{j,k}$ for the entanglement witness coming from
the density matrix corresponding to the maximum value of the concurrence in Fig.\ref{concurrence}.
As we can see, the elements with the most significant
weights (greater than 0.05) for measuring the witness, correspond to
total of five measurements: two separate measurements of
$\sigma^{z}$ in each polariton, and two joint measurements
$\sigma_1^{z}\otimes\sigma_2^{z}$ and $\sigma_1^{x}\otimes\sigma_2^{y}$.

 The values of $\zeta$ and $\xi_{1}$
used in Fig.\ref{concurrence} to get maximal entanglement would correspond to
$\Delta=10J$, $\kappa=10J$, $G=\gamma=0.01J$ and the pumping
coherent field  was also taken to have roughly a hundred photons. $J$
is tunable and depends on the coupling of the photonic modes between
neighboring cavities. Assuming this to be of the order of
$10^{10}Hz$, this would correspond to a cavity dissipation rate
$\kappa\approx 10^{11}Hz$ and a polaritonic decay rate
$\gamma\approx10^{8}Hz$. These correspond to 0.1 nanoseconds
lifetime of the cavity field and to ten nanoseconds for the
polaritonic excitations at the two ends, which are within the near
future in technologies like coupled toroidal microcavities and
coupled superconducting qubits \cite{toroid}.  Coupled defect
cavities in photonic crystals arrays are also fast approaching this
dissipation regime and are extremely suited in fabrication of
regular arrays of many coupled defect cavities interacting with
quantum dots \cite{noda}. In all technologies, an increase in $J$, in
coupling between the cavity modes, the requirements on the various
lifetimes of the polaritonic and photonic field modes can be
further reduced.

\section{Conclusion}

To summarize, this paper presents an example of entangling two qubits in the presence of
dissipation despite the fact that each qubit has a continuously decaying
state. The entanglement is not transient but stationary, and thereby
easy to verify
in an experiment, for which there is also a relevant witness. Though the
amount of entanglement is not maximal,
it is still very interesting as it is for a completely open system.
As opposed to the
typical case of, say, many-body systems or even the case
of two purely atomic qubits in a single cavity or extremely
close as to be able to directly interact, here there is the
added advantage that the entangled qubits are easily
individually accessible (being encoded in distinct atom-cavity
systems) for measurements. It is worthwhile to point out an existing
scheme to have steady state entanglement between entities in
distinct cavities entangles atoms \cite{parkins} (as opposed to polaritons) and
is much more intricate.

It is very interesting and counterintuitive that only a classical laser field driving the central
cavity/connecting fiber was necessary to entangle the polaritonic
qubits. A scheme feasible with current or near future technology and able to verify
polaritonic entanglement as the one we have suggested in this paper,
would be a significant first step towards the realization of the plethora
schemes to simulate many-body systems and quantum computation using
coupled cavities.
Moreover, the model would also deserve to deepen
counterintuitive properties of entanglement against noise (see e.g. \cite{stoch}).

\subsection*{Acknowledgments}
This work has been supported by QIP IRC (GR/S821176/01),  and the
European Union through the Integrated Projects SCALA (CT-015714).
 SB would like to thank the Engineering and Physical Sciences Research
Council (EPSRC) UK for an Advanced Research Fellowship the support
of the Royal Society and the Wolfson foundation. SM thanks SB for
hospitality at University College London. We would like to thank Y.
Yamamoto for pointing out that a fiber can replace the central
cavity in the three cavity system.

\end{document}